\documentclass[11pt]{article}

\usepackage[final]{acl}

\usepackage{times}
\usepackage{latexsym}
\usepackage[T1]{fontenc}
\usepackage[utf8]{inputenc}
\usepackage{microtype}
\usepackage{inconsolata}
\usepackage{graphicx}
\usepackage{dsfont}
\usepackage{amsmath,amsfonts}
\usepackage{booktabs}
\usepackage{multirow}
\usepackage{enumitem}
\usepackage{algorithm}
\usepackage{placeins}
\usepackage{algpseudocode}
\usepackage{xcolor}
\usepackage{subcaption}
\usepackage{tabularx}
\usepackage{colortbl}
\usepackage{url}
\usepackage{diagbox}
\usepackage{amssymb}
\definecolor{hlblue}{HTML}{2E75B6}
\definecolor{hlgreen}{HTML}{548235}
\definecolor{hlred}{HTML}{C0504D}
\definecolor{hlamber}{HTML}{BF8F00}

\newcommand{\high}{\textsc{High}}
\newcommand{\midd}{\textsc{Mid}}
\newcommand{\low}{\textsc{Low}}
\newcommand{\step}[1]{\textbf{Step~#1}}

\title{CORE: A Unified Cascaded Ordinal Relevance Estimation Framework for E-commerce Search}

\author{
  Zhi Jin\textsuperscript{1,*}, Xi Wang\textsuperscript{2,*}, Yunfei Li\textsuperscript{1}, Guojun Liu\textsuperscript{1}, Qingsong Hua\textsuperscript{1}, Wei Lin\textsuperscript{1,\dag} \\
  \textsuperscript{1}Meituan \quad \textsuperscript{2}Beijing Institute of Technology \\
  \{jinzhi07, liyunfei18, liuguojun08, huaqingsong, linwei31\}@meituan.com, xiwangai@bit.edu.cn
}

\begin{document}

\maketitle

\begingroup
\renewcommand{\thefootnote}{}
\footnotetext{\hspace{-1.8em}\textsuperscript{*}\,Equal contribution.}
\footnotetext{\hspace{-1.8em}\textsuperscript{\dag}\,Corresponding author.}
\endgroup

\begin{abstract}
Ranking relevance is a fundamental task in e-commerce search, directly affecting ranking quality and consumer experience. Although inherently an ordinal classification problem, it is commonly formulated as conventional multi-class classification, which overlooks the natural order among relevance levels and assigns equal penalties to adjacent and distant misclassifications. This mismatch leads to suboptimal learning objectives for practical relevance evaluation. To address this issue, we propose a unified cascaded binary classification framework applicable to both large language model inference and online BERT-based inference, which reformulates relevance estimation as a sequential decision process and decomposes multi-class prediction into a series of ordered binary judgments from higher to lower relevance tiers. For large language models, we design a step-wise reasoning procedure with pruning strategies and tier-specific reward functions. For the online BERT model, we replace the conventional classification head with multiple level-wise binary classifiers and distill the capabilities of large language models into the online model. Extensive offline industrial benchmark evaluations and online A/B experiments demonstrate that the proposed framework substantially improves relevance performance, reducing the online bad-case rate by 15.94\%. Further analyses suggest that tier-wise modeling is effective for relevance estimation.
\end{abstract}

\section{Introduction}
\label{sec:intro}

Relevance is a fundamental task in e-commerce search systems, responsible for assessing how well each candidate item matches a given user query. Typically, search relevance is modeled as an ordinal classification problem with multiple ordered levels, each governed by fine-grained domain-specific rules that vary across query intent types. For instance, an item may be labeled as highly relevant if it exactly matches the user's query intent, weakly relevant if it can serve as a substitute, and irrelevant if it is not substitutable or fails to satisfy the intent.

\begin{figure}[t]
  \centering
  \includegraphics[width=\columnwidth]{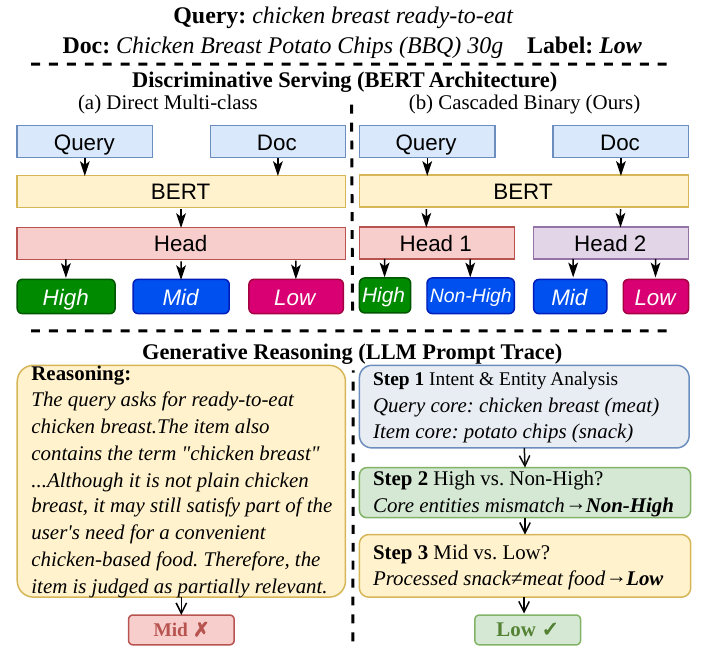}
  \caption{Comparison of direct multi-class and cascaded binary classification on BERT (upper) and LLM (lower).}
  \label{fig:cascaded-example}
\end{figure}

This fine-grained formulation poses substantial challenges for e-commerce relevance assessment, as models are required to capture user intent, reason over product attributes, and identify nuanced substitutability relations under domain-specific grading criteria. To address these challenges, pre-trained language models have been increasingly adopted for semantic relevance modeling, evolving from BERT-based encoders~\citep{nogueira2019passage} to sequence-to-sequence ranking models~\citep{nogueira2020document,zhuang2022rankt5}. More recently, large language models have further advanced this paradigm by enabling generative relevance assessment, in which models produce not only final relevance labels but also explicit reasoning traces that justify their decisions~\citep{thomas2024large,qin2024large,ma2023zero,pradeep2023rankzephyr,sun2023rankgpt}. Inspired by chain-of-thought reasoning~\citep{wei2022chain}, such reasoning-enhanced approaches improve both interpretability and grading accuracy by making the intermediate decision process more transparent~\citep{thomas2024large,zeng2025optimizing,mehrdad2024llm,dong2025taosr1}. Furthermore, reinforcement learning methods such as GRPO~\citep{shao2024deepseekmath} have recently emerged as a promising direction for aligning generative relevance models with task-specific reward signals, reflecting the broader trend of enhancing and aligning LLM reasoning capabilities through reinforcement learning~\citep{ouyang2022training,guo2025deepseek,zeng2025optimizing,dong2025taosr1}.

Despite these advances, most existing methods still treat relevance as a flat multi-class classification problem~\citep{reddy2022shopping,agrawal2025rationale}, directly assigning each query--item pair to a relevance category while neglecting the inherent ordinal structure. This formulation is misaligned with the nature of relevance assessment, where decision boundaries between adjacent levels are asymmetric: identifying highly relevant items often requires strict matching of core entities and user intent, whereas lower-level distinctions may depend on softer criteria such as partial satisfaction or substitutability. Treating ordered levels as independent classes therefore obscures such asymmetries and entangles heterogeneous reasoning criteria.

Motivated by this observation, we formulate relevance assessment as cascaded binary classification, decomposing ordinal prediction into ordered binary decisions. This design aligns with the ordinal structure of relevance labels, simplifies individual decisions, and enables step-wise verification. We instantiate the framework for both online classifiers and large language models: a cascaded two-head BERT architecture for low-latency serving, and cascaded reasoning with step-level GRPO for stronger LLM-based inference through fine-grained credit assignment.

Our main contributions are as follows:
\begin{itemize}[leftmargin=*,nosep,topsep=4pt]
  \item We identify the ordinal nature of e-commerce relevance assessment and propose a cascaded binary classification framework that better aligns model decisions with ordered relevance semantics.

  \item We instantiate this framework across both efficient online models and large language models, using cascaded BERT heads for low-latency deployment and step-level GRPO for fine-grained optimization of LLM reasoning.

  \item We conduct large-scale offline experiments and online A/B tests in a production e-commerce search engine, demonstrating consistent improvements over flat classification baselines and measurable reductions in relevance errors under real-world traffic.
\end{itemize}

\section{Related Work}
\label{sec:related}

\subsection{Search Relevance Models}

Pre-trained language models have become central to modern search relevance systems~\citep{lin2020pretrained}, enabling rankers to capture semantic matching beyond lexical overlap. BERT-based rankers~\citep{nogueira2019passage} and sequence-to-sequence models~\citep{nogueira2020document,zhuang2022rankt5} have achieved strong performance and broad industrial adoption~\citep{zou2021pre,yin2016ranking}. More recently, large language models have shifted relevance assessment toward prompting-based paradigms, including pointwise scoring~\citep{thomas2024large}, pairwise comparison~\citep{qin2024large}, and listwise reranking~\citep{ma2023zero,pradeep2023rankzephyr,sun2023rankgpt,zhu2023large}. Chain-of-thought reasoning has also been explored to improve interpretability and fine-grained relevance judgments~\citep{wei2022chain,kojima2022large,thomas2024large,zeng2025optimizing,mehrdad2024llm,dong2025taosr1,zhuang2023beyond}. Unlike concurrent work such as TaoSR1~\citep{dong2025taosr1} and GenCLS++~\citep{he2025gencls}, our work proposes a cascaded task decomposition framework that explicitly models asymmetric decision boundaries in e-commerce search and can be applied to both discriminative and generative models.

\subsection{Reinforcement Learning for LLM}

RLHF~\citep{ouyang2022training}, commonly implemented with PPO~\citep{schulman2017proximal}, has become a standard alignment paradigm, while DPO~\citep{rafailov2023direct} avoids explicit reward modeling. For reasoning-intensive tasks, GRPO~\citep{shao2024deepseekmath} has been adopted to encourage complex reasoning in mathematics, code generation~\citep{guo2025deepseek,yu2025dapo}, and more recently ranking-oriented reasoning~\citep{zhuang2025rankr1}. In search relevance, \citet{zeng2025optimizing} applied GRPO with Stepwise Advantage Masking for per-step credit assignment, but their method requires intermediate score generation and does not directly verify the correctness of individual reasoning steps. In contrast, our approach decomposes relevance assessment into semantically distinct decisions, each independently verifiable using its own binary ground-truth label.

\section{Methodology}
\label{sec:method}

Cascaded binary classification reformulates multi-way relevance prediction as ordered threshold-like decisions, aligning with the ordinal structure of relevance labels while enabling step-wise supervision and fine-grained credit assignment without an additional process reward model.

\subsection{Preliminaries}
\label{sec:method-prelim}

Let $x=(q,d)$ denote an input pair consisting of a query $q$ and a
candidate item $d$, and let $y^\star\in\{\high,\midd,\low\}$ be its gold
relevance label. The task is to predict $\hat{y}\in\{\high,\midd,\low\}$
from $x$. We consider two modeling paradigms: a \emph{discriminative}
approach, where an encoder $f_\theta$ (e.g., BERT) produces a
representation $h=f_\theta(x)$ followed by a classification head; and a
\emph{generative} approach, where a causal language model $\pi_\theta$
autoregressively produces a reasoning trace $t=(t_1,\ldots,t_T)$ from
which a prediction $\hat{y}$ is extracted. Our cascaded binary
classification framework applies to both paradigms, as detailed below.

\subsection{Cascaded Binary Classification}
\label{sec:method-cascade}

Cascaded binary classification reformulates multi-way relevance prediction as a sequence of ordered binary decisions rather than a single flat classification problem. The cascade first separates the most critical category, such as highly relevant items, from the remaining candidates, and then resolves finer-grained distinctions only when necessary. This formulation aligns with the asymmetric decision boundaries in relevance assessment and explicitly encodes the ordinal structure among labels. By decomposing the task into successive threshold-like decisions, each stage can focus on a more specialized criterion, reducing the entanglement of different reasoning patterns.

The cascaded formulation also enables step-wise verifiability. Given a gold label, the correct decision path is uniquely determined, allowing each active step to be directly supervised by comparing the predicted path with the gold path. This enables fine-grained credit assignment without requiring an additional process reward model. Steps skipped due to early termination contribute neither reward nor gradient, a property leveraged by the training strategies introduced below.

\subsection{Cascaded Classification for LLM Reasoning}
\label{sec:method-llm}

For large language models, we embed the cascaded binary decisions into a structured reasoning trace and train the model with supervised fine-tuning followed by step-level GRPO.

\subsubsection{\textbf{Cascaded Reasoning Schema}}
\label{sec:method-llm-schema}

\begin{figure}[t]
  \centering
  \includegraphics[width=\columnwidth]{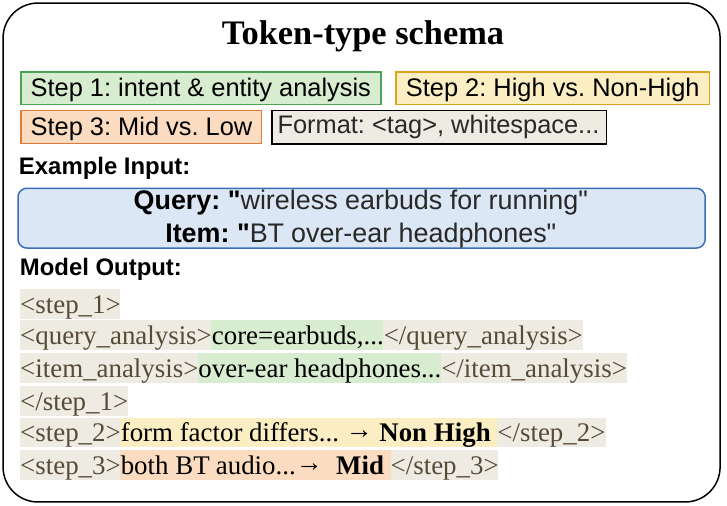}
  \caption{Token-type schema annotated on one reasoning trace. The trace is partitioned into four groups. Each group receives an independent learning signal during step-level RL training.}
  \label{fig:token-schema}
\end{figure}

For three-level e-commerce relevance assessment, we instantiate the cascade as a structured reasoning trace with three steps:
\begin{itemize}[leftmargin=*,nosep,topsep=2pt]
  \item \step{1} \emph{Intent and entity analysis}: identify the query intent, core entities, and modifiers in q, as well as the product category and key attributes of d.
  \item \step{2} \emph{High vs.\ non-high}: determine whether d satisfies the criteria for \high\ relevance. If so, output $\hat{r}=\high$ and terminate.
  \item \step{3} \emph{Mid vs.\ low}: if \step{2} predicts non-high, further decide whether d is partially relevant (\midd) or irrelevant (\low).
\end{itemize}
Unlike unconstrained free-form reasoning, this schema aligns the reasoning trajectory with an explicit sequence of classification decisions. Figure~\ref{fig:token-schema} illustrates the token-type schema on a concrete example.

\subsubsection{\textbf{SFT Warm-Up}}
\label{sec:method-sftwarmup}

Reinforcement learning is initialized from a fine-tuned model. To build this warm-up model,
we use the training set with gold relevance labels and generate multiple
reasoning traces for each query-item pair. Specifically, we sample ten
candidate traces for every training instance under the cascaded
reasoning schema described above, and retain only those samples whose
final predicted label matches the human annotation~\citep{wang2022self}. In this way, the
supervised training data is filtered by task correctness.

Formally, for each training instance $x_i$ with gold label $r_i^\star$,
we sample a set of candidate traces $\{y_i^{(m)}\}_{m=1}^{10}$ and
construct the warm-up set as
\begin{equation}
\mathcal{D}_{\mathrm{SFT}} =
\left\{
\big(x_i, y_i^{(m)}\big) \;\middle|\;
\hat{r}\!\left(y_i^{(m)}\right)=r_i^\star
\right\}.
\end{equation}

This filtering strategy ensures that the
initial policy learns a reasoning pattern that is consistent with the
final business label, while removing noisy samples lie in the training data.
We then perform supervised fine-tuning on the retained traces to obtain
the warm-up policy. This
stage provides a more stable starting point.

\subsubsection{\textbf{Step-Level GRPO}}
\label{sec:method-stepgrpo}

\begin{figure*}[t]
  \centering
  \includegraphics[width=0.98\textwidth]{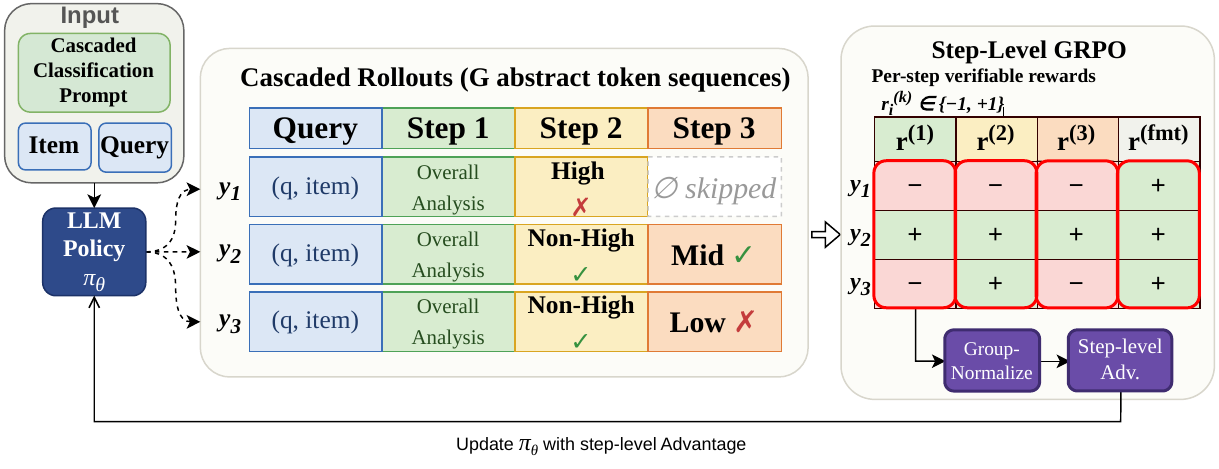}
  \caption{Illustration of step-level GRPO for cascaded reasoning. \textbf{Left:} the LLM policy $\pi_\theta$ samples $G$ cascaded rollouts for the same (query, item) pair, where Step~3 is skipped whenever Step~2 predicts High. \textbf{Right:} the step-level RL pipeline---per-step verifiable rewards $r_i^{(k)}\!\in\!\{-1,+1\}$ are computed against the gold label, group-normalized into step-level advantages, and broadcast as token-level credit for policy update via clipped GRPO with KL regularization.}
  \label{fig:main}
\end{figure*}

Standard GRPO normalizes a single scalar reward across the response group
and broadcasts it to every token. This is suboptimal for our setting because a response is not a
single homogeneous sequence but a cascaded one. Therefore ,we design a step-level RL training scheme
that decomposes reward computation, normalization, and step-level
credit assignment according to the structure of the reasoning chain, as illustrated in Figure~\ref{fig:main}.

\begin{table}[t]
\centering
\small
\begin{tabularx}{\columnwidth}{lX}
\toprule
Token group & Learning signal \\
\midrule
\step{1} & Correctness on the final answer \\
\step{2} & Correctness on the high versus non-high decision \\
\step{3} & Correctness on the mid versus low decision, active only on the non-high branch \\
Format & Correctness on structural format  \\
\bottomrule
\end{tabularx}
\caption{Token groups and their learning signals in step-level RL training.}
\label{tab:token-groups}
\end{table}

\paragraph{\textbf{Reward design.}}

For each sampled response $y_i$, we compute four scores
$s_i^{(1)}$, $s_i^{(2)}$, $s_i^{(3)}$, and
$s_i^{(\mathrm{fmt})}$, corresponding to the four token groups in
Table~\ref{tab:token-groups}. Let $\hat{r}_i$ denote the final predicted
label extracted from $y_i$, and let $r^\star$ denote the gold label.
The \step{1} score is tied to the correctness of the final answer:
\begin{equation}
s_i^{(1)} =
\begin{cases}
1, & \hat{r}_i = r^\star,\\
-1, & \text{otherwise}.
\end{cases}
\end{equation}
This treats the first-step analysis as useful if it supports a correct
final judgment and harmful otherwise.

For the binary decision steps, we define
$z^{(2)}=\mathds{1}[r^\star=\high]$ and
$z^{(3)}=\mathds{1}[r^\star=\midd]$ for non-high samples, where $\mathds{1}[\cdot]$ denotes the indicator function. Let
$b_i^{(2)}$ and $b_i^{(3)}$ denote the binary decisions made in
\step{2} and \step{3}, respectively. Their scores are
\begin{equation}
\begin{aligned}
s_i^{(2)} &=
\begin{cases}
1, & b_i^{(2)} = z^{(2)},\\
-1, & \text{otherwise},
\end{cases}
\\[4pt]
s_i^{(3)} &=
\begin{cases}
1, & b_i^{(3)} = z^{(3)},\\
-1, & b_i^{(3)} \neq z^{(3)},\\
0, & r^\star=\high.
\end{cases}
\end{aligned}
\end{equation}
The last case reflects the fact that \step{3} is inactive on the
\high\ branch and should not contribute a training signal. Finally,
$s_i^{(\mathrm{fmt})}$ measures whether the response preserves the
required structure. In our implementation, this score is derived from
basic format validity checks and is intentionally much weaker than the
semantic decision scores.

\paragraph{\textbf{Stepwise GRPO and token-level credit assignment.}}

Let $\mathcal{G}$ be the set of responses sampled for the same
query-item pair. Instead of normalizing a single response-level reward,
we normalize each score type independently. For each
$k\in\{1,2,3,\mathrm{fmt}\}$, we compute
\begin{equation}
\label{eq:stepwise_norm}
\begin{aligned}
\mu_k &= \mathrm{mean}_{j\in\mathcal{G}}\, s_j^{(k)}, \\
\sigma_k &= \mathrm{std}_{j\in\mathcal{G}}\, s_j^{(k)}, \\
A_i^{(k)} &= \frac{s_i^{(k)}-\mu_k}{\sigma_k+\epsilon}.
\end{aligned}
\end{equation}
This independent normalization is the key difference from standard
GRPO. Each group is compared only against responses at the same stage,
which prevents the easier steps from dominating the harder ones.

Let $g_{i,t}\in\{1,2,3,\mathrm{fmt}\}$ denote the step membership of
token $t$ in response $y_i$, determined by its position within the
structured trace. Each token directly inherits the advantage of its
step, with a downweighting factor $\omega_k$ for the format group:
\begin{equation}
\label{eq:token_adv}
A_{i,t} = \omega_{g_{i,t}}\,A_i^{(g_{i,t})},
\end{equation}
where $\omega_1=\omega_2=\omega_3=1$ and
$\omega_{\mathrm{fmt}}=\lambda_{\mathrm{fmt}}<1$. As a result, different
regions of the same response receive different learning signals: if
\step{2} is wrong but \step{1} is correct, the model is encouraged to
preserve the useful analysis in \step{1} while revising the decision
logic in \step{2}.

The final StepGRPO objective extends the clipped GRPO surrogate with
per-step token-level advantages and a KL penalty to the SFT reference
policy:

\begin{equation}
\label{eq:sgrpo_obj}
\small
\begin{aligned}
\mathcal{L}_{\mathrm{SGRPO}}(\theta)
= {} & -\mathbb{E}
\left[
\frac{1}{G}\sum_{i=1}^{G}
\bar{\ell}_i(\theta)
\right] \\
& + \beta\,\mathrm{KL}
\big(\pi_\theta \,\|\, \pi_{\mathrm{ref}}\big), \\
\bar{\ell}_i(\theta)
= {} & \frac{1}{|y_i|}
\sum_{t=1}^{|y_i|}
\ell_{i,t}(\theta), \\
\ell_{i,t}(\theta)
= {} & \min\left(
\rho_{i,t}A_{i,t},
\bar{\rho}_{i,t}A_{i,t}
\right).
\end{aligned}
\end{equation}

where the expectation is over prompts $x$ and sampled responses 
$\{y_i\}_{i=1}^{G}$. Here, $\rho_{i,t}$ is the per-token importance ratio, and 
$\bar{\rho}_{i,t}=\mathrm{clip}(\rho_{i,t},1-\epsilon,1+\epsilon)$ is its clipped 
version. StepGRPO provides finer-grained credit assignment along the reasoning 
chain via token-level optimization.

\begin{algorithm}[t]
\caption{Step-level RL training procedure}
\label{alg:step-level-rl}
\small
\begin{algorithmic}[1]
\Require training data $\mathcal{D}=\{(x,r^\star)\}$, current policy $\pi_\theta$, reference policy $\pi_{\mathrm{ref}}$, group size $G$, KL coefficient $\beta$, learning rate $\eta$
\Ensure updated policy $\pi_\theta$
\While{training not converged}
  \State Sample a minibatch $\{(x,r^\star)\}\sim\mathcal{D}$
  \For{each training instance $(x,r^\star)$}
    \State Generate a response group $\{y_i\}_{i=1}^{G}\sim\pi_\theta(\cdot\mid x)$
    \For{$i=1$ to $G$}
      \State Parse $y_i$ into \step{1}, \step{2}, \step{3}, and format token groups
      \State Extract final label $\hat{r}_i$ and binary decisions $b_i^{(2)}, b_i^{(3)}$
      \State Compute step scores $s_i^{(1)}, s_i^{(2)}, s_i^{(3)}, s_i^{(\mathrm{fmt})}$
    \EndFor
    \For{each $k\in\{1,2,3,\mathrm{fmt}\}$}
      \State Compute group statistics $\mu_k,\sigma_k$ and advantages $A_i^{(k)}$ using Eq.~\eqref{eq:stepwise_norm}
    \EndFor
    \For{$i=1$ to $G$}
      \For{each token $y_{i,t}$}
        \State Identify token group $g_{i,t}$ and mask $m_{i,t}$
        \State Assign token-level advantage $A_{i,t}$ using Eq.~\eqref{eq:token_adv}
      \EndFor
    \EndFor
  \EndFor
  \State Compute objective $\mathcal{L}_{\mathrm{StepGRPO}}(\theta)$ using Eq.~\eqref{eq:sgrpo_obj}
  \State Update $\theta \leftarrow \theta - \eta \nabla_\theta \mathcal{L}_{\mathrm{StepGRPO}}(\theta)$
\EndWhile
\State \Return $\pi_\theta$
\end{algorithmic}
\end{algorithm}

\subsection{Cascaded Classification for BERT}
\label{sec:method-bert}

The cascaded binary principle applies directly to lightweight
discriminative models deployed in online search. Given an input pair
$x=(q,d)$, a BERT encoder produces a pooled representation
$h=\text{BERT}(x)\in\mathbb{R}^d$. Instead of a flat multi-class head
mapping $h$ to $C$ logits, we apply two binary heads:

\begin{itemize}[leftmargin=*,nosep,topsep=2pt]
  \item \textbf{Head~1 (High vs.\ Non-High):} a binary classifier with
    parameters $(\mathbf{w}_1,b_1)$ that computes the probability of
    \high:
    \begin{equation}
    p_1 = \sigma(\mathbf{w}_1^\top h + b_1),
    \end{equation}
    where $\sigma$ is the sigmoid function. If $p_1 > \tau_1$ for a
    calibrated threshold $\tau_1$, the system emits \high\ and terminates.
  \item \textbf{Head~2 (Mid vs.\ Low):} invoked only when $p_1 \le \tau_1$,
    a second binary classifier with parameters $(\mathbf{w}_2,b_2)$
    computes:
    \begin{equation}
    p_2 = \sigma(\mathbf{w}_2^\top h + b_2).
    \end{equation}
    If $p_2 > \tau_2$, the system emits \midd; otherwise it emits \low.
\end{itemize}

The final prediction is:
\begin{equation}
\hat{y} =
\begin{cases}
\high, & p_1 > \tau_1,\\
\midd, & p_1 \le \tau_1 \land p_2 > \tau_2,\\
\low,  & \text{otherwise}.
\end{cases}
\end{equation}

Both heads share the same encoder and differ only in their projection
layers, introducing no additional inference cost. Let
$y_1=\mathds{1}[y^\star=\high]$ and $y_2=\mathds{1}[y^\star=\mathrm{Mid}]$ be
the binary targets for the two heads. Training minimizes:
\begin{equation}
\begin{aligned}
\mathcal{L}_{\text{cascade}} =
&\;\mathcal{L}_{\text{BCE}}(p_1,\,y_1) \\
&\;+\;(1-y_1)\cdot\mathcal{L}_{\text{BCE}}(p_2,\,y_2).
\end{aligned}
\label{eq:cascade_loss}
\end{equation}

\subsection{LLM-to-BERT Distillation via PostCoT}
\label{sec:method-distill}

The cascaded LLM described in Section~\ref{sec:method-llm} produces
high-quality reasoning traces but incurs generation latency that is
prohibitive for real-time online serving. To bridge this gap, we adopt a
PostCoT strategy inspired by TaoSR1~\citep{dong2025taosr1} and distill
the LLM's cascaded reasoning into the lightweight cascaded BERT.

First, we use the trained cascaded LLM to re-inference on the training
data, sampling reasoning trajectories for each query--item pair.
Each trajectory is reformatted into PostCoT style, where the predicted
label is placed before the step-by-step reasoning trace. We then perform
supervised fine-tuning of LLM on these
PostCoT-formatted traces, obtaining a PostCoT LLM that generates
label-first responses with cascaded reasoning.

Since the label appears at a fixed prefix position, we can directly
extract the LLM's output logits at first position. Let
$\mathbf{z} = [z_{\high}, z_{\midd}, z_{\low}]^\top$ denote the
3-class logits from the PostCoT LLM. To transfer this knowledge, we
align each binary head of the cascaded BERT with the corresponding
aggregation of the LLM's logits.

For \textbf{Head~1} (High vs.\ Non-High), we aggregate the non-high
logits into a single score via log-sum-exp, preserving the logit scale:
\begin{equation}
\mathbf{z}_{\text{LLM}}^{(1)} =
\big[\,\log(e^{z_{\midd}} + e^{z_{\low}}),\;\; z_{\high}\,\big]^\top.
\end{equation}
For \textbf{Head~2} (Mid vs.\ Low), we directly take the low and mid
logits as the conditional binary target:
\begin{equation}
\mathbf{z}_{\text{LLM}}^{(2)} = [z_{\low},\; z_{\midd}]^\top.
\end{equation}

Let $\mathbf{z}_{S1}, \mathbf{z}_{S2} \in \mathbb{R}^2$ denote the
cascaded BERT's two head outputs. The training objective extends the
cascaded BCE loss (Equation~\ref{eq:cascade_loss}) with a
KL-divergence distillation term~\citep{hinton2015distilling} for each head:
\begin{equation}
\begin{aligned}
\mathcal{L}_{1} = {} & \mathcal{L}_{\text{BCE}}(p_1, y_1) \\
& + \lambda \cdot T^2 \cdot \mathcal{L}_{\text{KL}} \big( \sigma_T(\mathbf{z}_{S1}), \sigma_T(\mathbf{z}_{\text{LLM}}^{(1)}) \big)
\end{aligned}
\end{equation}
\begin{multline}
\mathcal{L}_{2} = (1-y_1) \cdot \Big[
\mathcal{L}_{\text{BCE}}(p_2, y_2)
+ \lambda \cdot T^2 \\
\cdot \mathcal{L}_{\text{KL}}\big(
\sigma_T(\mathbf{z}_{S2}),
\sigma_T(\mathbf{z}_{\text{LLM}}^{(2)})
\big) \Big],
\end{multline}
where $y_1, y_2$ are the binary gold labels, $\sigma_T$ denotes softmax
with temperature $T$, and $\lambda$ controls the distillation weight.
The $(1-y_1)$ gating ensures Head~2 only receives learning signal on
non-high instances. The total objective is $\mathcal{L} =
\mathcal{L}_{1} + \mathcal{L}_{2}$, transferring the LLM's cascaded
reasoning into the dual-head BERT.

\begin{table*}[t]
\centering
\footnotesize
\setlength{\tabcolsep}{0pt}
\begin{tabular*}{\textwidth}{@{\extracolsep{\fill}}lcccccccccc@{}}
\toprule
\multirow{2}{*}{Method}
& \multicolumn{3}{c}{High}
& \multicolumn{3}{c}{Mid}
& \multicolumn{3}{c}{Low}
& \multirow{2}{*}{Acc.} \\
\cmidrule(lr){2-4}\cmidrule(lr){5-7}\cmidrule(lr){8-10}
& P & R & F1 & P & R & F1 & P & R & F1 & \\
\midrule
\multicolumn{11}{l}{\textit{LLM-based Methods}} \\
Direct-GRPO
& 0.7187 & 0.8567 & 0.7817
& 0.7241 & \textbf{0.6259} & 0.6714
& \textbf{0.8063} & 0.7609 & 0.7829
& 0.7478 \\
Cascaded-SFT
& 0.7032 & 0.8394 & 0.7653
& 0.7243 & 0.6034 & 0.6584
& 0.7781 & 0.7572 & 0.7675
& 0.7333 \\
Cascaded-GRPO
& \underline{0.7219} & \textbf{0.8665} & \underline{0.7876}
& \underline{0.7504} & 0.6171 & \underline{0.6772}
& \underline{0.7967} & \underline{0.7787} & \underline{0.7876}
& \underline{0.7541} \\
Cascaded-StepGRPO
& \textbf{0.7338} & \underline{0.8620} & \textbf{0.7928}
& \textbf{0.7832} & \underline{0.6201} & \textbf{0.6922}
& 0.7860 & \textbf{0.8123} & \textbf{0.7990}
& \textbf{0.7648} \\
\midrule
\multicolumn{11}{l}{\textit{LLM-PostCoT Methods}} \\
TaoSR1\citep{dong2025taosr1}
& \textbf{0.7505} & 0.8432 & 0.7941
& 0.7528 & \textbf{0.6591} & 0.7028
& 0.7834 & 0.7842 & 0.7838
& 0.7621 \\
PostCoT-CORE
& \underline{0.7344} & \textbf{0.8725} & \textbf{0.7975}
& \textbf{0.7990} & \underline{0.6183} & \textbf{0.6971}
& \textbf{0.7909} & \textbf{0.8210} & \textbf{0.8057}
& \textbf{0.7706} \\
\midrule
\multicolumn{11}{l}{\textit{BERT-based Methods}} \\
Direct-BERT
& 0.7322 & \underline{0.8281} & 0.7864
& 0.7414 & 0.6297 & 0.6857
& 0.7683 & 0.7744 & 0.7628
& 0.7441\\
Cascaded-BERT
& \underline{0.7336} & \textbf{0.8432} & \underline{0.7928}
& \textbf{0.7536} & \underline{0.6420} & \underline{0.6979}
& \underline{0.7821} & \underline{0.7824} & \underline{0.7717}
& \underline{0.7558} \\
Cascaded-BERT-Distilled
& \textbf{0.7505} & \textbf{0.8432} & \textbf{0.7941}
& \underline{0.7528} & \textbf{0.6591} & \textbf{0.7028}
& \textbf{0.7834} & \textbf{0.7842} & \textbf{0.7838}
& \textbf{0.7622} \\
\bottomrule
\end{tabular*}
\caption{Main results of both LLM-based and BERT-based methods on the e-commerce relevance benchmark.}
\label{tab:main-results}
\end{table*}

\section{Experiments}
\label{sec:experiments}

\subsection{Experimental Settings}

We evaluate the proposed cascaded binary classification framework under two complementary settings. First, in an offline evaluation, we compare LLM-based and BERT-based variants on a human-annotated relevance benchmark. Second, in an online evaluation, we deploy a cascaded BERT classifier in a production e-commerce search engine and assess its impact on user experience.

\subsubsection{\textbf{Benchmark and Metrics}}

We evaluate all methods on a large-scale human-annotated e-commerce relevance benchmark from a production search engine, containing 90{,}000 query--item pairs from 44{,}621 unique queries. This benchmark is selected to assess the effectiveness of the proposed method in realistic e-commerce search scenarios. Labels are uniformly distributed across three relevance levels: \high, \midd, and \low. We report accuracy (ACC) as the overall metric, along with per-class precision, recall, and F1-score.

\subsubsection{\textbf{Training Data}}

We construct the training data from real-world e-commerce search logs. Training data are constructed from real-world e-commerce search logs. We sample \textbf{1.2 million} annotated query--item pairs from \textbf{9 million} production instances for LLM SFT warm-up, while the full \textbf{9 million} instances are used for small-model distillation.

\subsubsection{\textbf{Implementation Details}}

All experiments are conducted on a distributed cluster with 8 NVIDIA A100 (80GB) GPUs. LLM training is implemented in PyTorch with \texttt{LlamaFactory}~\citep{zheng-etal-2024-llamafactory} for SFT and \texttt{veRL}~\citep{sheng2025verl} for reinforcement learning.

For SFT warm-up, we fine-tune Qwen3-14B for 2 epochs using AdamW ($\beta_1=0.9, \beta_2=0.95$, weight decay 0.1). The peak learning rate is $2\times10^{-5}$ with cosine decay and a 0.05 warmup ratio. The global batch size is 256, and the maximum sequence length is 4096.

For Step-Level GRPO, training starts from the SFT policy and runs for 1200 steps with \texttt{veRL}. The global batch size is 256, and the peak learning rate is $1\times10^{-6}$ under a cosine schedule. Each prompt generates $G=8$ rollouts via \texttt{vLLM} with temperature 0.7. We set the KL coefficient to $\beta=0.001$, clipping parameter to $\epsilon=0.2$, format reward weight to $\lambda_{\mathrm{fmt}}=0.1$, and context window to 4096 tokens.

For Cascaded-BERT, we adopt \texttt{bge-small-zh-v1.5} ~\citep{xiao2023bge} as the encoder and fine-tune it with cascaded classification heads for 3 epochs using AdamW ($\beta_1=0.9, \beta_2=0.999$, weight decay 0.01). The global batch size is 256, the maximum sequence length is 256, and the peak learning rate is $2\times10^{-5}$ with linear decay and a 0.05 warmup ratio.

For distillation, the student is trained with both hard labels and LLM soft logits, using $\lambda=0.5$ and $T=4.0$ in the conditional binary KL-divergence objective~\citep{hinton2015distilling}. During calibrated cascade inference, thresholds are set to $\tau_1=0.54$ and $\tau_2=0.50$ to balance precision and recall.

\subsection{Offline Evaluation}

\subsubsection{\textbf{Models}}

All LLM-based methods conducted use Qwen3-14B~\citep{yang2025qwen3} as the backbone.
Table~\ref{tab:main-results} compares seven systems below:

\begin{itemize}[leftmargin=*,nosep,topsep=2pt]
  \item \textbf{Direct-GRPO}: SFT followed by response-level GRPO on the direct three-class formulation, used to assess the effect of cascaded modeling.

  \item \textbf{Cascaded-SFT}: a supervised warm-up model trained on filtered high-quality cascaded reasoning traces.

  \item \textbf{Cascaded-GRPO}: Cascaded-SFT further optimized with response-level GRPO using a single sequence-level reward.

  \item \textbf{Cascaded-StepGRPO}: our full LLM-based model, applying step-level GRPO with normalized per-step rewards and token-level advantage broadcasting.

  \item \textbf{TaoSR1}: an LLM-based Taobao Search relevance framework that leverages the PostCoT strategy. We adopt it as a representative LLM-based relevance modeling approach and evaluate it on our dataset.

  \item \textbf{PostCoT-CORE}: a TaoSR1-style PostCoT LLM trained on cascade-classification rationales.

  \item \textbf{Direct-BERT}: a pre-trained BERT model fine-tuned with cross-entropy loss for direct three-class classification.

  \item \textbf{Cascaded-BERT}: a BERT-based cascaded model with two sequential binary classification heads.

  \item \textbf{Cascaded-BERT-Distilled}: a Cascaded-BERT model distilled from Cascaded-StepGRPO using the PostCoT method in Section~\ref{sec:method-distill}.
\end{itemize}

\subsubsection{\textbf{Overall Performance}}

Table~\ref{tab:main-results} summarizes the main results.

\textbf{Cascaded reasoning improves over direct classification.}
Cascaded-GRPO improves the overall accuracy over Direct-GRPO from 0.7478 to 0.7541, with consistent F1 gains across all three relevance levels. This suggests that decomposing relevance assessment into sequential binary decisions better captures the ordered decision boundaries among relevance classes than direct multi-class classification.

\textbf{Step-level credit assignment further enhances GRPO.}
Cascaded-StepGRPO achieves the best performance among LLM-based methods, reaching an accuracy of 0.7648 and outperforming Cascaded-GRPO by 1.07 percentage points. It also obtains the highest F1 scores across all relevance levels within this group, indicating that step-level rewards provide more fine-grained optimization signals than response-level GRPO.

\textbf{PostCoT reasoning further strengthens LLM-based relevance prediction.}
LLM-PostCoT methods achieve additional gains over standard LLM-based approaches. TaoSR1 reaches an accuracy of 0.7621, while PostCoT-CORE further improves it to 0.7706, achieving the best overall performance among all LLM-based variants. PostCoT-CORE also obtains consistently strong F1 scores across High, Mid, and Low relevance levels, suggesting that post-hoc chain-of-thought supervision helps the model better align with the latent decision process of relevance assessment.

\textbf{PostCoT-CORE improves boundary-sensitive decisions.}
Compared with TaoSR1, PostCoT-CORE improves accuracy by 0.85 percentage points and yields consistent F1 gains across all relevance levels, especially for Low relevance. This indicates that CORE-style PostCoT reasoning is effective in reducing confusion near relevance decision boundaries and improving prediction balance across classes.

\begin{figure}[t]
  \centering
  \includegraphics[width=\columnwidth]{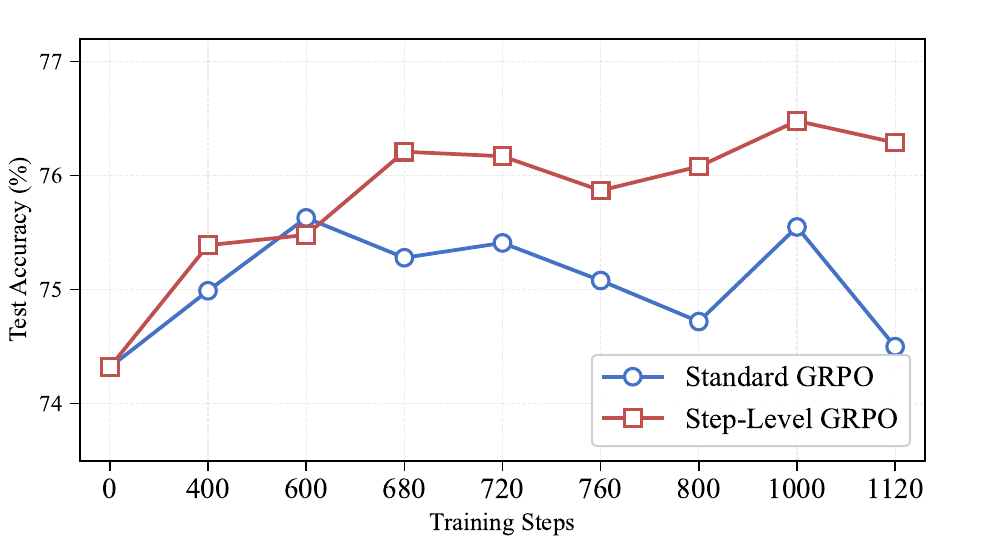}
  \caption{Test accuracy of Standard GRPO and Step-Level GRPO on the e-commerce benchmark.}
  \label{fig:grpo-comparison}
\end{figure}

\begin{figure}[t]
  \centering
  \includegraphics[width=\columnwidth]{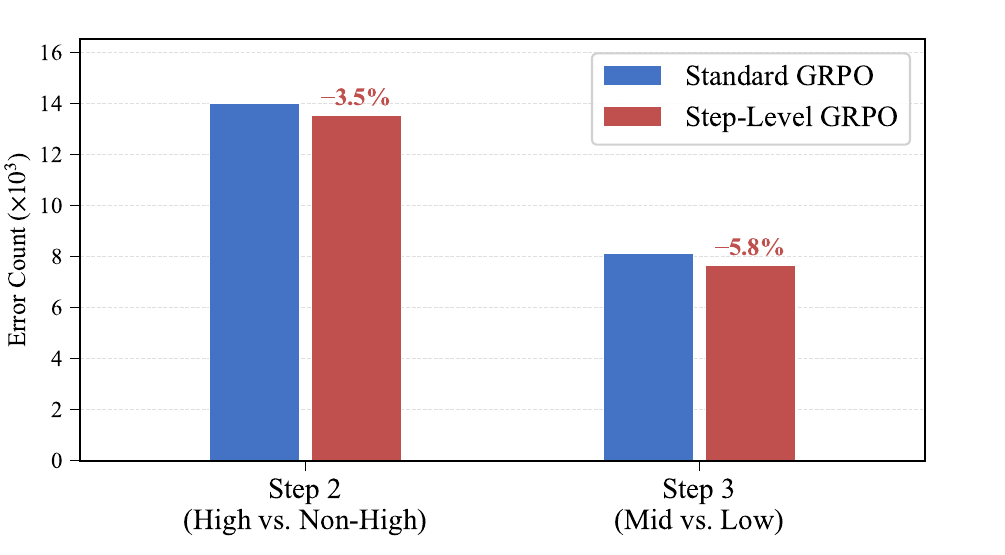}
  \caption{Per-step decision error counts on the e-commerce benchmark.}
  \label{fig:step-errors}
\end{figure}

\textbf{Step-Level GRPO exhibits more stable convergence.}
Figure~\ref{fig:grpo-comparison} shows that Step-Level GRPO is initialized from the same Cascaded-SFT checkpoint as Standard GRPO and begins to outperform it after Step~400, demonstrating improved training stability and optimization efficiency. Figure~\ref{fig:step-errors} further shows that step-level credit assignment reduces errors at both decision boundaries, suggesting that it improves the two binary decisions jointly rather than benefiting only a single stage.

\textbf{The cascaded structure also benefits BERT-based models.}
Cascaded-BERT improves accuracy over Direct-BERT by 1.17 percentage points and achieves consistent F1 gains across all relevance levels. This confirms that the cascaded structure is also effective for lightweight discriminative models, making it suitable for online deployment scenarios.

\textbf{Distillation further improves cascaded BERT.}
Cascaded-BERT-Distilled improves accuracy over Cascaded-BERT from 0.7558 to 0.7622 and achieves the best F1 scores among BERT-based methods. This demonstrates that transferring the LLM's cascaded reasoning ability into the dual-head BERT through PostCoT distillation provides complementary gains beyond the structural benefit of the cascaded design.

\subsection{Online Evaluation}

To satisfy the strict latency requirements of production serving, we deploy the Cascaded-BERT architecture (Section~\ref{sec:method-bert}) for online inference, rather than an LLM-based model. For online deployment, we set the confidence thresholds $\tau_1$ and $\tau_2$ to 0.54 and 0.50, respectively. We conduct an A/B test on our production experimentation platform, comparing Cascaded-BERT with the existing flat multi-class BERT baseline.

We evaluate online performance using two top-5 metrics: \textbf{NDCG@5} and \textbf{Badcase@5}. NDCG@5 measures position-aware ranking quality under graded relevance, while Badcase@5 measures the fraction of queries for which at least one top-5 result contains a severe relevance error.

As shown in Table~\ref{tab:online-ab}, Cascaded-BERT improves NDCG@5 from 0.9020 to 0.9038 ($+0.20\%$) and reduces Badcase@5 from 13.8\% to 11.6\%, a 15.9\% relative reduction. These results indicate that the cascaded architecture not only improves top-ranked relevance quality but also substantially reduces severe relevance failures. This is consistent with its explicit modeling of ordered decision boundaries, which discourages overestimating low-relevance items into high-ranked positions.

\begin{table}[htb]
\centering
\small
\begin{tabularx}{\columnwidth}{l>{\centering\arraybackslash}X>{\centering\arraybackslash}X}
\toprule
Method & NDCG@5 & Badcase@5 \\
\midrule
Baseline & 0.9020 & 13.8\% \\
Ours & 0.9038 & 11.6\% \\
\bottomrule
\end{tabularx}
\caption{Online A/B test results}
\label{tab:online-ab}
\end{table}

\section{Conclusion}
\label{sec:conclusion}

In this paper, we introduced a cascaded binary classification framework for e-commerce search relevance estimation, an inherently ordinal classification task. The framework decomposes relevance estimation into a sequence of conditional binary decisions, thereby reducing the difficulty of direct multi-class optimization while explicitly capturing the ordinal structure of relevance labels. We demonstrated its generality across LLM-based generative reasoning and BERT-based discriminative inference, and validated its effectiveness through large-scale offline experiments and online A/B testing on an industrial e-commerce search platform. We hope this work provides a practical foundation for applying cascaded binary decomposition to broader ordinal classification problems.

\bibliography{references}

\end{document}